\documentclass[pra,showpacs,preprintnumbers,amsmath,amssymb,superscriptaddress]{revtex4}

\usepackage{bm}
\usepackage{graphicx}
\usepackage{amsbsy}
\usepackage{amsmath}
\usepackage{amsfonts}
\usepackage{amsthm}

\begin{document}

\theoremstyle{plain}
\newtheorem{theorem}{Theorem}
\newtheorem{lemma}[theorem]{Lemma}
\newtheorem{corollary}[theorem]{Corollary}
\newtheorem{conjecture}[theorem]{Conjecture}
\newtheorem{proposition}[theorem]{Proposition}

\theoremstyle{definition}
\newtheorem{definition}{Definition}

\theoremstyle{remark}
\newtheorem*{remark}{Remark}
\newtheorem{example}{Example}

\title{How many ebits can be unlocked with one classical bit?}   
\author{Gilad Gour}\email{gour@math.ucalgary.ca}
\affiliation{Institute for Quantum Information Science and 
Department of Mathematics and Statistics,
University of Calgary, 2500 University Drive NW,
Calgary, Alberta, Canada T2N 1N4} 

\date{\today}

\begin{abstract}
We find an upper bound on the rate at which entanglement can be unlocked
by classical bits. In particular, we show that for quantum information sources
that are specified by ensambles of pure bipartite states, one classical bit can 
unlock at most one ebit.
\end{abstract}  


\maketitle

Both classical and quantum correlations can be locked in quantum 
states~\cite{Coh98,DiV98,DiV04,Hor04,Chr05}.
The idea that classical information can unlock the entanglement that is hidden in a
quantum state was first introduced in~\cite{Coh98,DiV98}. Later on, 
it has been shown~cite{Hor04} that there exist measures of entanglement that are 
\emph{lockable} in the sense that they can decrease arbitrarily after measuring one 
qubit. In particular, it has been found that the entanglement of formation, entanglement
cost, logarithmic negativity, and recently the squashed entanglement~\cite{Chr05} are all lockable 
measures, whereas the relative entropy of entanglement is a non-lockable measure. 

In this paper we view lockable measures from the opposite direction. That is,
instead of considering the loss of entanglement subject to discarding or measurement 
of one qubit, we consider the gain in the entanglement shared by two parties 
(Alice and Bob) after receiving one classical bit from a third 
party~\footnote{This view is more similar to the one given in~\cite{Coh98}.}. 
This view is equivalent to the one introduced in~\cite{Hor04} and in fact
one can easily construct an example similar to the one given in~\cite{Hor04} for which
instead of measuring one qubit in order to decrease entanglement, a third party send
Alice and Bob one classical bit and as a result increase arbitrarily their shared 
entanglement of formation. Viewing it in this direction helps us to define the rate at which entanglement can be unlocked with classical bits.

Despite the fact that it is possible (in some cases) to increase arbitrarily the entanglement of formation (and some other measures) with one classical bit, it is still an open important question weather it is possible to unlock arbitrarily large number of singlets with one classical bit. This question is related to the question whether the distillable entanglement is a lockable
measure or not. To my knowledge the answer to this later question is unknown, although 
as we will show below, if the the quantum information source is specified by an ensemble
of pure bipartite states, one classical bit can unlock at most one ebit. 
   
Consider an i.i.d. quantum source, $\mathcal{S}^\text{AB}$, that is specified by an ensemble        
$\{p_i, \sigma_i\}$ of \emph{bipartite} quantum states, and that consecutive uses of the source are independent and produce the state $\sigma_{i}$ with probability $p_i$. 
In particular, $N$ consecutive uses of the source produce the state 
$\sigma_{i_1}\otimes\sigma_{i_2}\otimes\cdots\otimes\sigma_{i_N}$ with probability
$p_{i_1}p_{i_2}...p_{i_N}$.  Given a bipartite measure of entanglement, $E$, we define the entanglement of  $\mathcal{S}^\text{AB}$ as:
\begin{equation}
E\left(\mathcal{S}^\text{AB}\right)\equiv\lim_{N\rightarrow\infty}\frac{1}{N}
\left\langle E\left(\sigma_{i_1}\otimes\sigma_{i_2}\otimes\cdots\otimes\sigma_{i_N}\right)\right\rangle\;,
\end{equation}
where $\langle \cdots \rangle$ denotes an average over all the possible states
$\sigma_{i_1}\otimes\sigma_{i_2}\otimes\cdots\otimes\sigma_{i_N}$.
Note that if $E$ is an additive measure of entanglement then
$$
E\left(\mathcal{S}^\text{AB}\right)=\sum_{i}p_iE\left(\sigma_i\right)\;.
$$
Suppose now that after $N$ consecutive uses of the source the supplier
distributes the bipartite state $\sigma_{i_1}\otimes\cdots\otimes\sigma_{i_N}$ to Alice
and Bob without informing them about the values of $i_1$, $i_2$, etc. We assume, 
however, that Alice and Bob know the statistics of the source $\mathcal{S}^\text{AB}$  
and therefore, from their perspective, they end up sharing the state $\rho^{\otimes N}$,
with $\rho\equiv\sum_{i}p_i\sigma_i$. Hence, without the classical information about
the values of $i_1$,...,$i_N$, the average number of ebits (per one use of the source) 
shared between Alice and Bob is give by:
$$
E^{\infty}(\rho)=\lim_{N\rightarrow\infty}\frac{1}{N}E\left(\rho^{\otimes N}\right)\;.
$$
The difference $E\left(\mathcal{S}^\text{AB}\right)-E^{\infty}(\rho)$ is therefore
the maximum possible increment (per copy) in entanglement due to additional 
classical information. Hence, given a bipartite quantum information source 
$\mathcal{S}^\text{AB}$, and a bipartite measure of entanglement $E$,
the maximum rate at which ebits (measured with $E$) can be unlocked by 
calssical bits is given by
\begin{equation}
\mathcal{R}_{E}\left(\mathcal{S}^\text{AB}\right)
\equiv\frac{E\left(\mathcal{S}^\text{AB}\right)-E^{\infty}(\rho)}{H(\{p_i\})}\;,
\end{equation}
where $H(\{p_i\})$ is the Shannon entropy of the distribution $\{p_i\}$. 
In this paper we will assume that $E$ is a proper measure of entanglement;
that is, $E$ is an entanglement monotone which is equal to the entropy of entanglement
on pure states and which is also asymptotically continuous.  
We point out 
that one can also define the \emph{locking capacity}, $\mathcal{L}_\varepsilon$,
of a (bipartite) quantum channel $\varepsilon$ as
$$
\mathcal{L}_\varepsilon\equiv\max_{\mathcal{S}^\text{AB}}
\mathcal{R}_{E}\left(\varepsilon\left(\mathcal{S}^\text{AB}\right)\right)\;,
$$
where $\varepsilon\left(\mathcal{S}^\text{AB}\right)\equiv\{p_i,\;\varepsilon(\sigma_i)\}$
and the maximum is taken over all possible quantum information sources.

\begin{theorem}
Let $\mathcal{S}^\text{AB}=\{p_i,\;|\psi_{i}\rangle\}$ be an i.i.d. quantum information 
source that is specified by an ensemble of pure bipartite quantum states.
Then, for any proper measure of entanglement, $E$, the maximum number of ebits
that can be unlocked by one classical bit is bounded by 
$$
\mathcal{R}_{E}\left(\mathcal{S}^\text{AB}\right)\leq 1-
\frac{\left|S\left(\rho^\text{A}\right)-S\left(\rho^\text{B}\right)\right|}
{S\left(\rho^\text{AB}\right)}\;,
$$
where $\rho^\text{AB}\equiv\sum_{i}p_i |\psi_{i}\rangle\langle\psi_{i}|$ and $S(\cdot)$ is the von-Neumann entropy.
\end{theorem}
Note that $\mathcal{R}_{E}$ is always smaller than one and it is zero whenever
$S\left(\rho^\text{AB}\right)=\left|S\left(\rho^\text{A}\right)-S\left(\rho^\text{B}\right)\right|$.
It is an open question whether $\mathcal{R}_{E}$ can be zero for quantum information
sources with 
$S\left(\rho^\text{AB}\right)>\left|S\left(\rho^\text{A}\right)-S\left(\rho^\text{B}\right)\right|$.

In the following proof, we will make use of some properties of the von-Neumann entropy. 
In particular, the von-Neumann entropy satisfies
$$
0 \leq S\left(\sum_i p_i\sigma_{i}\right)-\sum_{i} p_i S\left(\sigma_{i}\right)
\leq H\left(\{p_i\}\right)\;,
$$
which also implies that for pure decompositions $\sigma_i=|\psi_{i}\rangle\langle\psi_{i}|$
we have
$$
S\left(\rho^\text{AB}\right)\leq H\left(\{p_i\}\right).
$$

\begin{proof}
For pure states, any proper measure of entanglement equals to the entropy of 
entanglement which is additive. Thus, 
$$
E(\mathcal{S}^\text{AB})=\sum_{i}p_iE\left(|\psi_i\rangle\right)\;,
$$
and $E(|\psi_{i}\rangle)=S(\rho^{A}_{i})=S(\rho^{B}_{i})$, where 
$\rho^{A}_{i}=\text{Tr}_\text{A}|\psi_{i}\rangle\langle\psi_{i}|$ is the reduced density 
matrix. From the concavity of the von-Neumann entropy we have:
\begin{equation}
\sum_{i}p_iE\left(|\psi_{i}\rangle\right)
\leq\min\Big\{S\left(\rho^\text{A}\right),\;S\left(\rho^\text{B}\right)\Big\}\;.
\label{up}
\end{equation}
The inequality above is usually strict
although in~\cite{Smo05} it has been shown that the regularized version of the entanglement 
of assistance equals
$\min\{S\left(\rho^\text{A}\right),\;S\left(\rho^\text{B}\right)\}$.
Now, since the distillable entanglement, $D$, provides a lower bound on any proper
measure of entanglement, we find a lower bound for $E^\infty (\rho^\text{AB})$
using the hashing inequality~\cite{Win05}. The hashing inequality provides a lower bound 
on the 1-way distillable entanglement
$$
D_{A\rightarrow B}\left(\rho^\text{AB}\right)\geq S\left(\rho^\text{A}\right)-S\left(\rho^\text{AB}\right)\;,
$$
where $D_{A\rightarrow B}$ is the 1-way distillable entanglement. Similarly, we have a lower bound for $D_{B\rightarrow A}$ which leads to
$$
D\left(\rho^\text{AB}\right)\geq\max
\Big\{S\left(\rho^\text{A}\right),\;S\left(\rho^\text{B}\right)\Big\}-S\left(\rho^\text{AB}\right)\;.
$$
Now, from Eq.~(\ref{up}) and the fact that 
$E^\infty\left(\rho^\text{AB}\right)\geq D\left(\rho^\text{AB}\right)$ we have
\begin{align}
& E\left(\mathcal{S}^\text{AB}\right)-E^\infty\left(\rho^\text{AB}\right)
\leq\; S\left(\rho^\text{AB}\right)\nonumber\\
&-
\left[\max\Big\{S\left(\rho^\text{A}\right),\;S\left(\rho^\text{B}\right)\Big\}
-\min\Big\{S\left(\rho^\text{A}\right),\;S\left(\rho^\text{B}\right)\Big\}
\right]\nonumber\\
& = S\left(\rho^\text{AB}\right)-\left|S\left(\rho^\text{A}\right)-S\left(\rho^\text{B}\right)\right|\;.
\end{align}
Hence, since $S\left(\rho^\text{AB}\right)\leq H\left(\{p_i\}\right)$ we get the upper bound
given in the theorem.
\end{proof}

The theorem above can be trivially generalized to the case when the quantum information
source is specified by an ensemble of pure multipartite states and the measure of
entanglement is taken to be the localizable entanglement~\cite{Pop05} or 
the entanglement of collaboration~\cite{Gou06}. 
This is due to the following two facts: (i) the localizable entanglement 
(or the entanglement of collaboration) satisfies Eq.~(\ref{up}) and (ii) the distillable
entanglement provides a lower bound for the localizable entanglement.

For quantum information sources that are specified by ensembles of mixed states
it is much more complicated to find an upper bound for $\mathcal{R}_{E}$ in general, 
and from the results in~\cite{Hor04} it can be arbitrarily large (i.e. depending on the
dimension or size of the Hilbert space). It is an interesting question whether
$\mathcal{R}_{D}\leq 1$, where $D$ is the distillable entanglement.

\emph{Acknowledgments:---}
I would like to thank Patrick Hayden and Aram Harrow for fruitful
discussions.

\end{document}